# Synthesis and characterization of $Bi_2Sr_2CaCu_2O_8$ ceramics prepared in presence of sodium

S. Rahier [a*], S. Stassen [a] , R. Cloots [a] and M. Ausloos [b]

[a] *Chemistry Institute B6, SUPRATECS, University of Liège, Sart-Tilman, B-4000 Liège, Belgium*

[b] *Physics Institute B5, SUPRATECS, University of Liège, Sart-Tilman, B-4000 Liège, Belgium*

**Abstract**

This paper aims to study Na doping in Bi-2212 superconducting ceramics. Na-doped Bi-2212 samples powders with a $Bi_{2-u}Sr_{2-v}Ca_{1-w}Cu_{2-x}Na_yO_z$ stoichiometry are prepared by means of thermal treatments. The intended Na substitution level, y, is fixed at 0.3. The initial content of the other ions is adopted so that Na substitution on one crystallographic site only is expected. Samples are characterized by electrical properties, X-ray diffraction analysis (XRD) and Energy Dispersive X-ray analysis (EDX). The influence of Na on the crystal chemistry and phases compatibility for 2212 materials is emphasized. Electrical resistivity results show that the critical temperature is independent of the sample initial stoichiometry. In some cases the production of a Bi-2223 phase is however enhanced.

## 1. Introduction

Since superconductivity with a high critical temperature ($T_c$) was found in the Bi-Sr-Ca-Cu-O system [1], these compounds have been extensively investigated. Bi-based high-temperature superconducting phases, represented by $Bi_2Sr_2Ca_{n-1}Cu_nO_{2n+4}$ (n = 1, 2 or 3) and written 2201, 2212 and 2223, are the first three members of the series. The 2212 phase is the most familiar one. Bi-2201 attracts less interest because of its low critical temperature (around 20 K) while the 2223 phase is intriguing because it cannot be easily prepared as a single-phase product, though it has a $T_c$ = 110 K.

The superconducting critical transition temperature of $Bi_2Sr_2CaCu_2O_{8+\delta}$ can be modified among other methods by the addition or substitution of elements of varying ionic radii and bonding characteristics. This variation is thought to be related to the density of charge carriers in the CuO planes. A strong modification would arise if an element with a free valence different from that of Bi, Sr, Ca or Cu is inserted in the lattice. However, since it is necessary to introduce the dopant into the system, a size constraint exists. Thus sodium metal seems to be a good candidate as a non isovalent dopant because its ionic radius $r_i$ (0.99-1.18 Å) depending on the configuration overlaps those of Bi, Sr, and Ca as illustrated in Table 1. Therefore, the charge-carrier concentration may be affected by such a substitution, hence a modification of physical properties related to the charge-carrier density can occur. More precisely, Na would be a good candidate to occupy the Bi place as far as the ionic size is concerned. On the other hand, a huge modification in the valence state, i.e., from +3 to +1 would arise.

Indeed, the substitution of Na for Bi and Ca would cause an electron redistribution among various layers, hence rendering the BiO and Ca planes less attractive to electrons. The consequent negative charge transfer to the SrO and CuO planes would reduce the hole density in $CuO_2$ planes to an optimum value, thus leading to an increase in $T_c$. The substitution of di- or trivalent ions by Na decreased the positive charge in the corresponding layer and thus reduced the oxygen content in order to fulfill the charge balance. In addition, another bonus can be gained from a technical point of view : the synthesis of 2212 materials with monovalent Na ion substitution for divalent cations is expected to occur at a temperature lower than the one needed in absence of alkali dopant, because the reaction rate in the solid state of the mixture containing Na ions is supposed to increase due to improved diffusion. Two arguments may be given to support the idea of a faster reaction in the presence of sodium carbonate. First, the melting temperature is around 600 °C, giving a liquid phase which is able to react much more rapidly than a solid. Secondly, it is possible that the mobility of the monovalent ion is much greater than that of divalent ones because of its lower charge and ionic size. Whence after the possible disappearance of the liquid phase by reaction with the other constituents, the process is governed by a solid state diffusion mechanism, controlled by the ion mobility. Thus, as for alkali-doped YBCO materials [3, 4], it is of great interest to study the effects of Na ion substitution in 2212 materials since beside providing some information on the chemical and physical properties of the system, Na doping might lead to an improvement of the technological properties, such as the superconducting critical temperature and the current carrying capacity.

*Corresponding author. Tel. : +32-4-366-3417 fax : + 32-4-366-3413
E-mail address : s.rahier@ulg.ac.be (S. Rahier)

Some authors report that Na is effective for producing $T_c$ enhancement. The work of Dou *et al.* [5] is devoted to the mechanism of $T_c$ enhancement by sodium doping in $Bi_{2.2}Sr_{1.8}Ca_{1.05}Cu_{2.15-y}Na_yO_{8+\delta}$ samples. A superconducting critical transition temperature of 94 K and a $T_{c\ off}$ (the temperature at which R(T) = 0) of 90 K has been achieved. Dou *et al.* observed that the *a*- and *c*-axes increased while the *b*-axis decreased with increasing the Na content, indirectly indicating that Na might have been introduced into the lattice even though it was not found on which crystallographic site Na was located.

The research of Yu *et al.* [6] was focused on the Na-doping of 2212 materials. They described the preparation of $Bi_{2.1}Sr_{1.8}Ca_{1.05}Na_yCu_{2.15-y}O_{8+\delta}$ (y = 0, 0.4, 0.5, 0.6, 0.7, 0.8, 1.0) samples using melt processing followed by quenching. The highest $T_c$ achieved was 94 K. During annealing in flowing oxygen, $T_c$ did not change. But $T_c$ was degraded under high oxygen pressure annealing. NMR analysis revealed that Na might have replaced Bi and Ca atoms.

It is questionable whether Na remains in the lattice. The question was also raised for the Na-YBCO case, see in [7-8] and refs. therein. Whence to start with a large Na concentration does not seem strictly relevant. Rather than fixing the Bi, Sr, Ca and Cu content near 2212 and attempting to introduce some Na into the lattice, it seems at first of interest to fix the Na concentration and let the ratios of Bi, Sr, Ca and Cu vary in order to observe whether various phases appear. From the two previous reports [5, 6] it seems that a y = 30 % Na concentration is of interest. Continuing preliminary studies [9], Na-doped Bi-2212 samples powders with a $Bi_{2-u}Sr_{2-v}Ca_{1-w}Cu_{2-x}Na_yO_{8+\delta}$ stoichiometry have been prepared, with y = 0.3, the initial content of the other ions is adopted so that Na substitution on one crystallographic site only is expected. The samples have been characterized by electrical measurements, XRD and EDX in order to analyse the effects of Na doping on the crystal chemistry of the 2212 phase. In the next section, experimental details are given, the third section is devoted to results. Section 4 contains a brief discussion on phase compatibility. Finally, concluding remarks are presented in the last section.

## 2. Experimental details

A stoichiometric 2212 phase powder has been prepared by mixing appropriate amounts of $Bi_2O_3$, $SrCO_3$, $CaCO_3$, $CuCO_3.Cu(OH)_2$, and $NaHCO_3$. The mixture has been treated at 820 °C for 48 h, with two intermediate grindings at regular intervals and a heating rate of 150 °C/h. The prereacted powder has been thereafter then mixed homogeneously with the necessary ingredients to reach a $Bi_{2-u}Sr_{2-v}Ca_{1-w}Cu_{2-x}Na_yO_{8+\delta}$ stoichiometry and pressed into a pellet. The substitution level has been fixed to y = 0.3. The pellet has been sintered at 750 °C with a heating rate of 150 °C/h during 24 h.

Room-temperature X-ray diffraction analysis, electric resistivity measurement, and EDX analysis have been performed on each pellet. The pellet was then put back in the furnace to be treated at 820 °C with a heating rate of 150 °C/h for 48 h. In Table 2, the 5 samples so studied are defined.

## 3. Results

The X-ray diffraction patterns for the $Bi_{2-u}Sr_{2-v}Ca_{1-w}Cu_{2-x}Na_{0.3}O_{8+\delta}$ compounds synthesized at 750 °C and then placed back in the furnace for further synthesis at 820 °C, respectively are given in Figs. 1 and 2. Besides the identified peaks corresponding to the 2212 structure, small amounts of 2201, CuO, $Bi_9Sr_{11}Ca_5O_z$, and $(Sr,Ca)_{14}Cu_{24}O_{41}$ phases were observed, together or not, depending on the initial stoichiometry, i.e., on which crystallographic site Na was supposed to be introduced. The 2201 phase, present in most samples treated at 750 °C (Fig. 1), decreased or even disappeared when the sintering temperature was raised to 820 °C (Fig. 2). EDX analyses were performed on each sample and are reported in Table 3. The first set of lines in Table 3 for each sample gives the composition at several spots after the heat treatment at 750 °C; the second set of lines gives the composition after the subsequent heat treatment at 820 °C at several other spots. The bolded composition reported in Table 3 is a mean value. The asterisk sign (*) indicates phases which were not found by EDX analysis likely because of the limited number of spots probed, though the corresponding phases were observed by X-ray diffraction analyses.

Electrical resistivity measurements were performed on each B1 to B5 sample by using the conventional four probe technique. The data are reported in Fig. 3 and characteristic temperatures are given in Table 4 for the samples (after treated at 820 °C). $T_{c\ on}$ and $T_{c\ off}$ are respectively the temperature at the beginning of the transition and the temperature at which the resistance vanishes. The resistivity transition width, $\Delta T_c$ was around 15 K in each sample, which indicates a distribution of superconducting grains with slightly different critical temperatures. The absence of foot structure below the inflection point at $T_c$ is indicative of good grain connectivity. It is remarkable that a jump is observed in the resistivity curve near 100 K. This indicated us the presence of a small amount of the 2223 high-Tc phase. This phenomenon is better marked in the B1 sample than in others.

**4. Discussion**

The presence of different phases in equilibrium with the 2212 phase can be correlated with a particular chemical composition for the 2212 phase obtained by the reaction : a strontium deficient chemical composition when the 2201 phase is in equilibrium with the 2212 phase, and a copper deficient chemical composition when CuO is in equilibrium with the 2212 phase. No trace of sodium ions was found in the 2212 structure by EDX. Moreover, the superconducting properties were independent on the chemical composition of the 2212 phases. In fact, the overall chemical composition of the intercalating layers (or "charge reservoir" layers) does not seem to change very much with the different approaches of doping : the bismuth content was around 2.2, i.e., in excess as compared to the ideal 2212 chemical composition; and the "strontium + calcium" content was around 2.8. Moreover, it is difficult to give any information concerning the oxygen concentration of the different phases: the presence of some 9<u>11</u>50 phase indeed modifies the thermodynamic equilibrium at 1 atmosphere oxygen pressure between the composition of the existing phases concerning the oxygen content.

Another important aspect of this study was the presence of a small amount of the 2223 phase in these materials, only visible in their electrical properties (see the small transition at -110 K in Fig. 3). The presence of the 2223 phase could be attributed to a driving force produced by the presence of sodium ions in the bulk, leading to the formation of the 2223 phase in a easy way as compared to a classical solid state route from stoichiometric chemical compositions. The thermodynamic stability domain of the 2212 and 2223 phases is thus strongly modified by the addition of a small amount of sodium in the initial composition.

**5. Conclusions**

We have prepared Na-doped Bi-2212 samples powders with a $Bi_{2-u}Sr_{2-v}Ca_{1-w}Cu_{2-x}Na_yO_{8+\delta}$ stoichiometry and analysed the obtained results by means of electrical measurements, XRD and EDX.
As it can be observed in Table 3, each 2212 superconducting phase is characterized by a bismuth concentration which is in excess as compared to the ideal chemical composition. Finally, from the electrical resistivity measurements as a function of temperature (as reported in Fig. 3), the bismuth substitution for strontium, and the strontium substitution for calcium (and conversely) seem to have no influence on the superconducting critical temperature for the synthesized 2212 samples.

## 6. Acknowledgments

Thanks to the MIEL laboratory (University of Liège) for electrical measurements.

## List of captions

Table 1.
Valence states and ionic radii for ions in 2212 materials substituted by sodium metal [2].

Table 2.
The initial chemical composition of the studied 2212 pellets with $Bi_{2-u}Sr_{2-v}Ca_{1-w}Cu_{2-x}Na_yO_{8+\delta}$ formula for the u, v, w, x, and y parameter values as given

Figure 1.
X-ray diffraction patterns for the $Bi_{2-u}Sr_{2-v}Ca_{1-w}Cu_{2-x}Na_{0.3}O_{8+\delta}$ samples as defined in Table 2 synthesized at 750 °C

Figure 2.
X-ray diffraction patterns for the $Bi_{2-u}Sr_{2-v}Ca_{1-w}Cu_{2-x}Na_{0.3}O_{8+\delta}$ samples as defined in Table 2 synthesized at 750 °C and further sintered at 820°C

Table 3.
Phases observed in the B1 to B5 Na-doped samples by EDX analysis and X-ray diffraction data

Figure 3.
Normalized electrical resistivity (ρ) versus temperature curves for the $Bi_{2-u}Sr_{2-v}Ca_{1-w}Cu_{2-x}Na_{0.3}O_{8+\delta}$ samples synthesized at 750°C and further sintered at 820°C

Table 4.
Electrical resistivity behavior.
* The average composition given is the one found at 750 °C, that at 820 °C was not available. Nevertheless, we can reasonably accept this value because the average composition for both sintering temperatures for the other samples differ only slightly.

## Illustrations

| | Valence state | Coordination | Ionic radius |
|---|---|---|---|
| Bi | +3 | 6 | 1.03 |
| Sr | +2 | 8/9 | 1.26/1.31 |
| Ca | +2 | 8 | 1.12 |
| Cu | +2 | 4 sq (2223) | 0.57 |
| | | 5 (2223-2212) | 0.65 |
| | | 6 (2201) | 0.73 |
| Na | +1 | 4 | 0.99 |
| | | 6 | 1.02 |
| | | 8 | 1.18 |

Table 1

| Na | u=v=w=x=0 | v=w=x=0 u=0.3 (-Bi) | u=w=x=0 v=0.3 (-Sr) | u=v=x=0 w=0.3 (-Ca) | u=v=w=0 x=0.3 (-Cu) |
|---|---|---|---|---|---|
| y=0.3 | B1 2212+Na | B2 2212-Bi+Na | B3 2212-Sr+Na | B4 2212-Ca+Na | B5 2212-Cu+Na |

Table 2

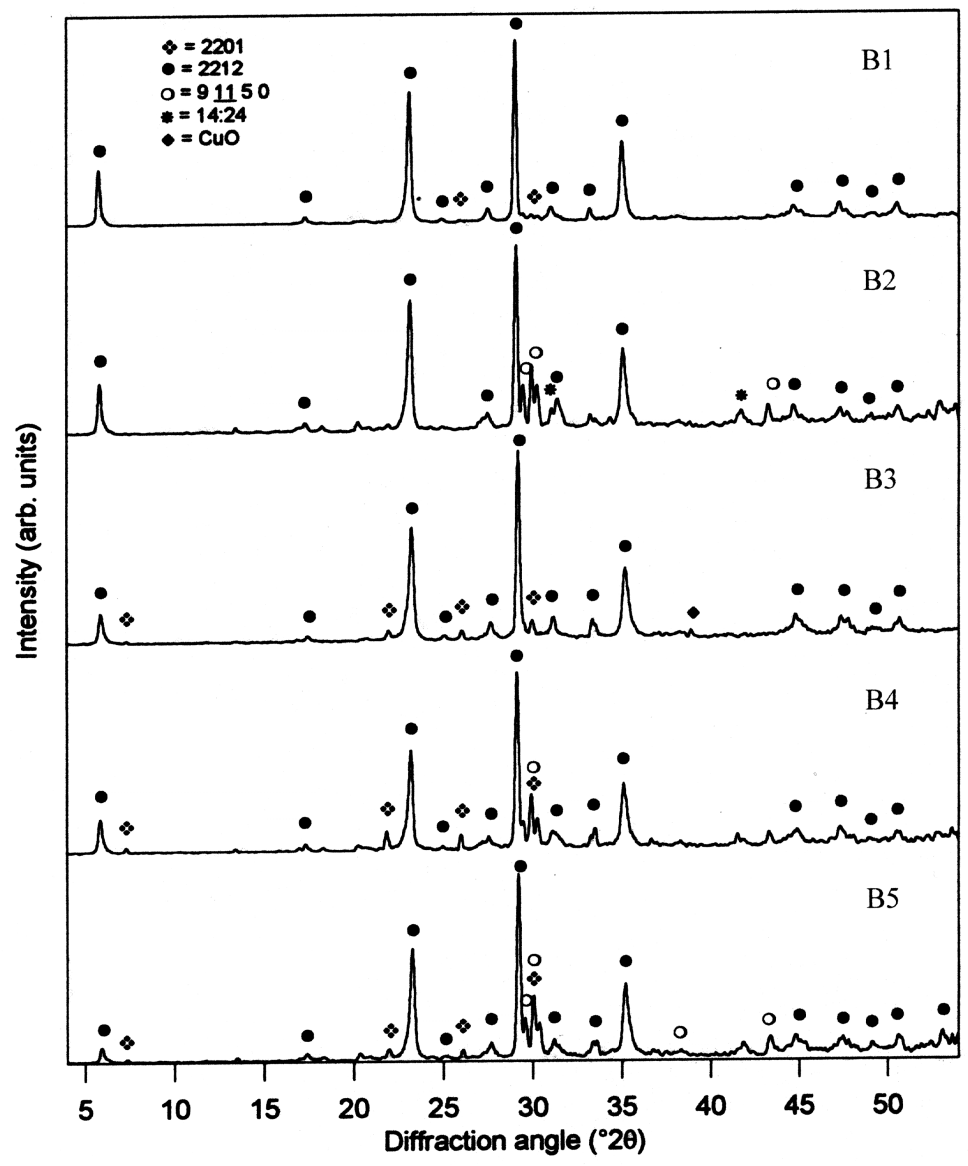

Figure 1

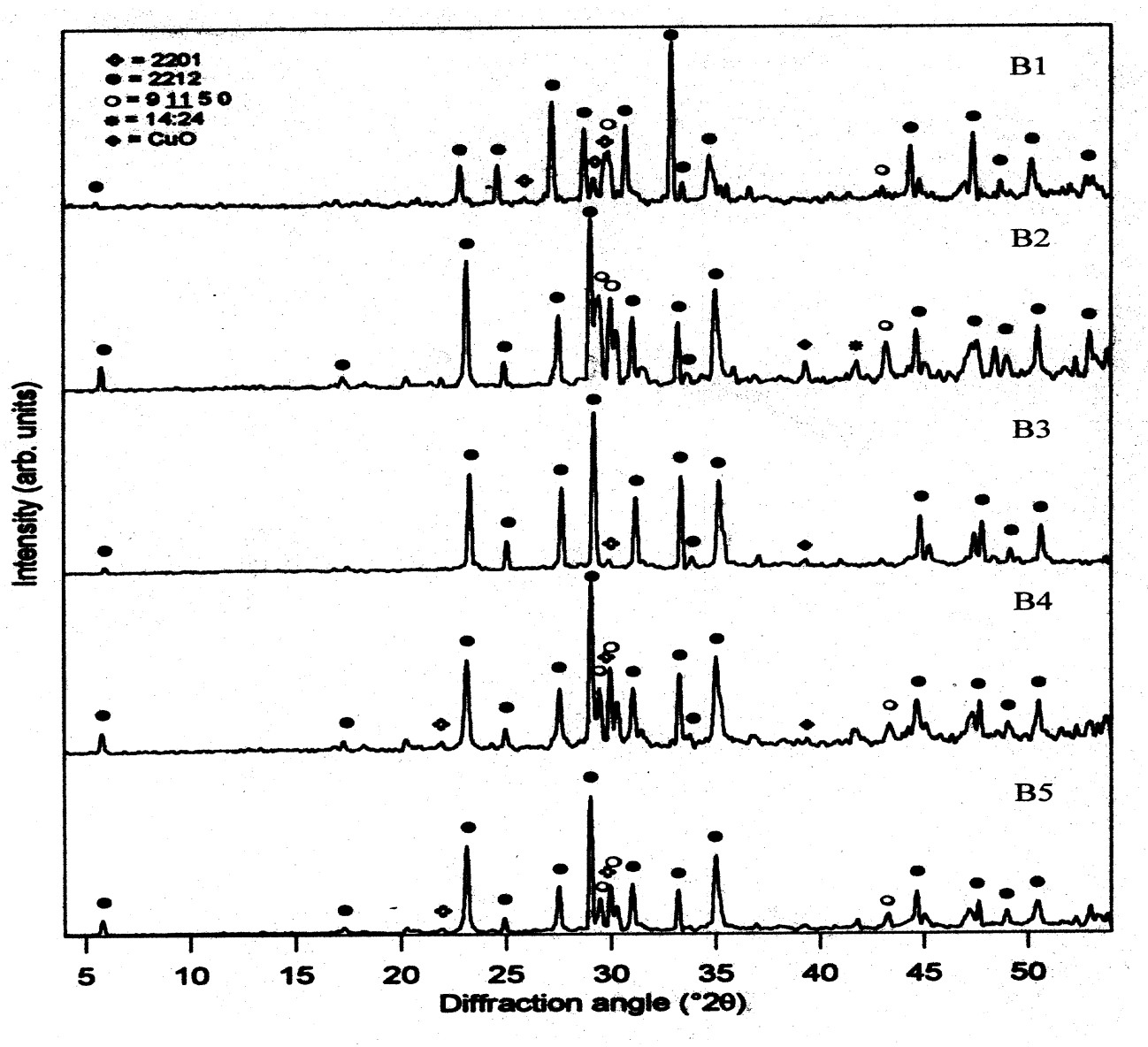

Figure 2

| # and initial stoichiometry | 2212 or like-phase | 2201 | 9 11 5 0 | CuO | other phases |
|---|---|---|---|---|---|
| B1<br>Na addition<br>$Bi_2Sr_2Ca_1Cu_2$ | $Bi_{2.15}Sr_{2.07}Ca_{0.77}Cu_{1.89}$<br>$Bi_{2.30}Sr_{2.03}Ca_{0.67}Cu_{1.57}$<br>$\mathbf{Bi_{2.23}Sr_{2.05}Ca_{0.72}Cu_{1.73}}$ | (*) | $Bi_9Sr_{11.96}Ca_{4.04}$ | - | - |
| | $Bi_{2.20}Sr_{1.94}Ca_{0.86}Cu_{1.78}$<br>$Bi_{2.18}Sr_{1.97}Ca_{0.84}Cu_{1.78}$<br>$\mathbf{Bi_{2.19}Sr_{1.96}Ca_{0.85}Cu_{1.78}}$ | (*) | $Bi_9Sr_{9.76}Ca_{4.23}$ | - | - |
| B2<br>Na on Bi-sites<br>$Bi_{1.7}Sr_2Ca_1Cu_2$ | $Bi_{2.12}Sr_{1.68}Ca_{1.18}Cu_{1.73}$<br>$Bi_{2.05}Sr_{1.65}Ca_{1.30}Cu_{2.16}$<br>$\mathbf{Bi_{2.08}Sr_{1.67}Ca_{1.24}Cu_{1.95}}$ | - | $Bi_9Sr_{10.68}Ca_{4.34}$ | CuO | $Sr_{10.47}Ca_{7.47}$<br>$Cu_{24}$ |
| | $Bi_{2.14}Sr_{2.02}Ca_{0.88}Cu_{1.83}$<br>$Bi_{2.20}Sr_{1.86}Ca_{0.96}Cu_{1.60}$<br>$\mathbf{Bi_{2.17}Sr_{1.94}Ca_{0.92}Cu_{1.72}}$ | - | $Bi_9Sr_{11.03}Ca_{4.82}$ | CuO | (*) |
| B3<br>Na on Sr-sites<br>$Bi_2Sr_{1.7}Ca_1Cu_2$ | | (*) | | | |
| | $Bi_{2.14}Sr_{1.90}Ca_{0.97}Cu_{1.73}$<br>$Bi_{2.23}Sr_{1.83}Ca_{0.97}Cu_{1.88}$<br>$Bi_{2.16}Sr_{1.98}Ca_{0.76}Cu_{1.78}$<br>$\mathbf{Bi_{2.18}Sr_{1.89}Ca_{0.91}Cu_{1.80}}$ | - | - | CuO | - |
| B4<br>Na on Ca-sites<br>$Bi_2Sr_2Ca_{1.7}Cu_2$ | $Bi_{2.23}Sr_{1.63}Ca_{1.09}Cu_{2.14}$<br>$Bi_{2.21}Sr_{1.56}Ca_{1.18}Cu_{2.18}$<br>$\mathbf{Bi_{2.22}Sr_{1.59}Ca_{1.13}Cu_{2.16}}$ | (*) | (*) | - | |
| | | (*) | (*) | (*) | |
| B5<br>Na on Cu-sites<br>$Bi_2Sr_2Ca_1Cu_{1.7}$ | $Bi_{2.13}Sr_{2.07}Ca_{0.83}Cu_{1.79}$<br>$Bi_{2.14}Sr_{1.98}Ca_{0.87}Cu_{1.90}$<br>$Bi_{2.18}Sr_{1.93}Ca_{0.94}Cu_{2.08}$<br>$\mathbf{Bi_{2.15}Sr_{1.99}Ca_{0.88}Cu_{1.92}}$ | (*) | $Bi_9Sr_{11.42}Ca_{4.18}$ | - | |
| | $Bi_{2.23}Sr_{1.68}Ca_{1.15}Cu_{1.95}$<br>$Bi_{2.16}Sr_{1.68}Ca_{1.04}Cu_{2.20}$<br>$\mathbf{Bi_{2.19}Sr_{1.68}Ca_{1.09}Cu_{2.07}}$ | (*) | $Bi_9Sr_{9.77}Ca_{4.94}$ | - | $Bi_{2.00}Sr_{1.20}$-<br>$Ca_{0.93}$ |

Table 3

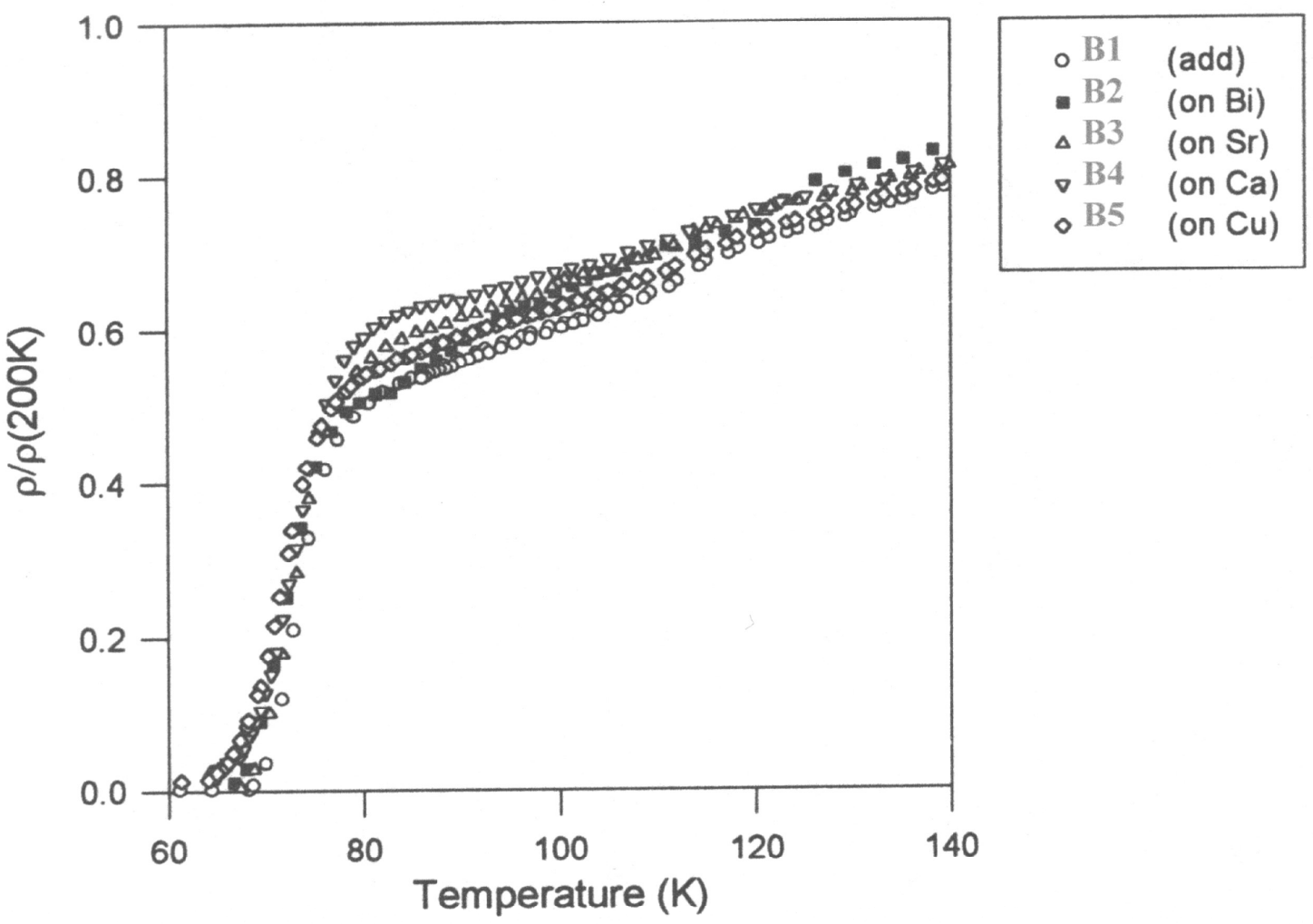

Figure 3

| # | Initial stoichiometry | Average composition | $T_{c\ on}$ | $T_{c\ off}$ | $T_c$ | ΔTc |
|---|---|---|---|---|---|---|
| B1 | $Bi_2Sr_2Ca_1Cu_2Na_{0.3}$ | $Bi_{2.19}Sr_{1.96}Ca_{0.85}Cu_{1.76}$ | 79 K | 66 K | 74 K | 13 K |
| B2 | $Bi_{1.7}Na_{0.3}Sr_2Ca_1Cu_2$ | $Bi_{2.17}Sr_{1.94}Ca_{0.92}Cu_{1.72}$ | 80 K | 67 K | 73 K | 13 K |
| B3 | $Bi_2Sr_{1.7}Na_{0.3}Ca_1Cu_2$ | $Bi_{2.18}Sr_{1.89}Ca_{0.91}Cu_{1.80}$ | 80 K | 68 K | 75 K | 12 K |
| B4 | $Bi_2Sr_2Ca_{0.7}Na_{0.3}Cu_2$ | $Bi_{2.22}Sr_{1.59}Ca_{1.13}Cu_{2.16}$ * | 83 K | 68 K | 74 K | 15 K |
| B5 | $Bi_2Sr_2Ca_1Cu_{1.7}Na_{0.3}$ | $Bi_{2.19}Sr_{1.68}Ca_{1.09}Cu_{2.07}$ | 79 K | 66 K | 74 K | 13 K |

Table 4